%% file: main.tex
\documentclass[runningheads]{llncs}
\usepackage[T1]{fontenc}
\usepackage{graphicx}
\usepackage[left]{lineno}
\usepackage{authblk}
\usepackage{array}
\usepackage{url,wrapfig}
\usepackage[misc]{ifsym}
\usepackage{booktabs}
\usepackage{array}
\usepackage{booktabs}
\usepackage{colortbl} 
\newcommand{\corr}{(\Letter)}
\usepackage{graphicx}
\usepackage{cite} 
\usepackage[linesnumbered,ruled,vlined,noend]{algorithm2e}  

\usepackage[table,xcdraw]{xcolor} 
\usepackage{amsmath}
\usepackage{amsmath, graphicx, xcolor, colortbl} 
\usepackage[font=small,skip=5pt]{caption} 
\usepackage{setspace} 

\usepackage{algpseudocode}
\usepackage{siunitx}
\usepackage{caption}
\usepackage{subcaption}

\makeatletter
\newcommand{\printfnsymbol}[1]{\textsuperscript{\@fnsymbol{#1}}}
\makeatother

\usepackage{hyperref} 
\usepackage{fontawesome}
\usepackage{mwe}
\makeatletter
\let\@@citation@@=\citation

\renewcommand{\citation}[1]{\@@citation@@{#1}%
\@for\@tempa:=#1\do{\@ifundefined{cit@\@tempa}%
  {\global\@namedef{cit@\@tempa}{}}{}}%
}

\def\@lbibitem[#1]#2#3\par{%
  \@ifundefined{cit@#2}{}{\item[\@biblabel{#1}\hfill]}%
  \if@filesw
      {\let\protect\noexpand
       \immediate
       \write\@auxout{\string\bibcite{#2}{#1}}}\fi\ignorespaces
  \@ifundefined{cit@#2}{}{#3}}
\def\@bibitem#1#2\par{%
  \@ifundefined{cit@#1}{}{\item}%
  \if@filesw \immediate\write\@auxout
    {\string\bibcite{#1}{\the\value{\@listctr}}}\fi\ignorespaces
  \@ifundefined{cit@#1}{}{#2}}
\makeatother

\begin{document}

\title{FastER: On-Demand Entity Resolution in Property Graphs}

\author{
Shujing Wang\inst{1}\thanks{These authors contributed equally to this work.} \and
Sibo Zhao\inst{1}\printfnsymbol{1} \and
Shiqi Miao\inst{1} \and
Selasi Kwashie\inst{2} \and
Michael Bewong\inst{3} \and
Junwei Hu\inst{1} \and
Vincent M. Nofong\inst{4} \and
Zaiwen Feng \inst{1,5,6}\corr
}

\institute{
College of Informatics, Huazhong Agricultural University, China \\
\email{zaiwen.feng@mail.hzau.edu.cn}\and
AI \& Cyber Futures Institute, Charles Sturt University, Australia \\
\and
School of Computing, Mathematics and Engineering, Charles Sturt University, Australia \\
\and
Department of Computer Science and Engineering, University of Mines and Technology, Ghana \\
\and
Hubei Key Laboratory of Agricultural Bioinformatics, China \\
\and
Engineering Research Center of Agricultural Intelligent Technology, Ministry of Education, China \\
}

\authorrunning{S. Wang et al.}
\titlerunning{FastER in Property Graphs}

\maketitle              

\begin{abstract}

Entity resolution (ER) is the problem of identifying and linking database records that refer to the same real-world entity. Traditional ER methods use batch processing, which becomes impractical with growing data volumes due to high computational costs and a lack of real-time capabilities. In many applications, users need to resolve entities for only a small portion of their data, making full data processing unnecessary---a scenario known as ``ER-on-demand''.
This paper proposes FastER, an efficient ER-on-demand framework for property graphs. Our approach uses {\em graph differential dependencies} (GDD) as a knowledge-encoding language to design effective filtering mechanisms that leverage both the structural and attribute semantics of graphs. We construct a blocking graph from the filtered subgraphs to reduce the number of candidate entity pairs that require comparison. Additionally, FastER incorporates Progressive Profile Scheduling (PPS), allowing the system to incrementally produce results throughout the resolution process.
Extensive evaluations on multiple benchmark datasets demonstrate that FastER significantly outperforms state-of-the-art ER methods in computational efficiency and real-time processing for on-demand tasks, without compromising quality or reliability.
We make FastER publicly available at the Github link~\href{https://github.com/Zaiwen/On_Demand_Entity_Resolution}{here}.

\end{abstract}

\section{Introduction}
Entity Resolution (ER) is the problem of determining when different data records represent 
the same real-world entity. While critical in domains like finance, healthcare, and social networks, ER faces significant computational challenges in 
\begin{wrapfigure}[15]{r}{0.55\textwidth}
\centering
\includegraphics[width=0.55\textwidth]{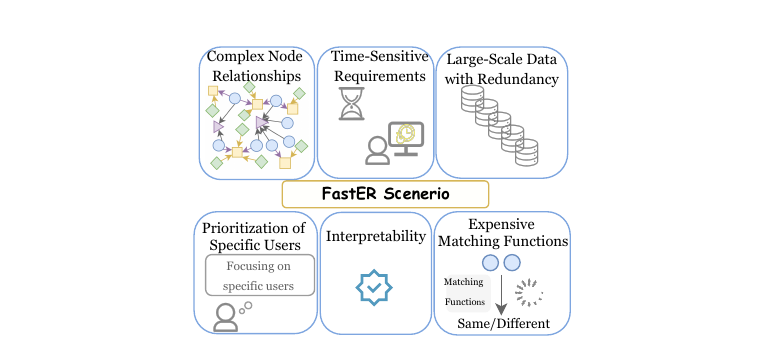}
\vspace{-9pt}
\caption{On-Demand ER in Property Graphs}
\label{fig:sce}
\end{wrapfigure}
big data. Indeed, its quadratic time complexity 
means every possible entity pair must be compared in the worst case. Traditional solutions like 
blocking~\cite{Christen2012a,Christen2012,Papadakis2022,Brinkmann2023}  and nearest-neighbor~\cite{Mann2016,Jiang2014,Wang2022} 
techniques reduce comparisons by filtering unlikely matches~\cite{Papadakis2023}. However, these methods often struggle 
with graph data, where entity relationships provide crucial context for identification. 
For example, in social networks and knowledge graphs, understanding connections among 
entities is essential for accurate duplicate detection. Moreover, conventional ER 
approaches may not meet the performance requirements of real-time applications where 
immediate results are needed. 
This paper studies the on-demand ER problem in property graphs, illustrated
in Example~\ref{ex:oer} below. 

\begin{example}[On-Demand ER Scenario]\label{ex:oer}
Consider a video streaming platform modeled as a property graph, where users, devices, and videos are represented as nodes with rich attributes, and complex relationships exist among them. Fig.~\ref{fig:sce} outlines six key requirements for entity resolution (ER) in this scenario.
First, batch-based ER methods are inefficient, especially when dealing with large-scale data, as they require processing the entire dataset before returning results. This makes it difficult to meet the responsiveness needed in time-sensitive applications and to prioritize relevant entities (Requirements~1, 2, and 4). Matching functions, such as neural or knowledge-based models, are often computationally expensive (Requirement 6), making pre-filtering essential for reducing unnecessary invocations. Moreover, interpretability is a crucial requirement in many real-world ER applications (Requirement~5).
Second, progressive ER methods struggle with the structural complexity of property graphs (Requirement~1) and are unable to focus on application-specific subgraphs (Requirement~4).
Existing on-demand ER solutions~\cite{BrewER, QDA, QueryER} are primarily designed for relational data and are not well-suited to the structural needs of graph-based scenarios (Requirement~1).
We propose FastER, an on-demand ER framework for property graphs that satisfies all six requirements outlined in Fig.~\ref{fig:sce}. \qed
\end{example}

\subsubsection{Contributions.} 
We present a graph differential dependencies (GDDs)~\cite{GDD} based framework for
on-demand entity resolution in property graphs.
Our approach dynamically processes relevant subgraphs 
and progressively emits cleaned entities, enabling direct querying of dirty property graphs. 
More formally, the contributions of the paper are summarised as follows: 1) We leverage GDDs 
to encode users' application-relevant domain knowledge, enabling relevance filtering  that utilizes both structural relationships and attribute semantics.   
2) Further, we propose an efficient and effective blocking solution for ER-on-demand tasks 
that reduces computational overhead while maintaining recall. 
Our approach extends traditional rule-based filtering with a targeted blocking scheme and a progressive scheduling scheme, representing the first adaptation of on-demand ER to graph data.
3) We conduct extensive experiments on benchmark graphs and relational datasets. The 
empirical results show that our method achieves significant improvements in efficiency, 
offering superior real-time processing capabilities and reliability compared
to existing solutions in ER-on-demand tasks.

\subsubsection{Related Work.} 
Works related to this paper fall under three groupings below. 

\noindent\underline{\em Graph Constraints.} In recent years, graph constraints have emerged 
as a prominent research area due to their relevance in data quality management. For instance, 
graph keys (GKeys)~\cite{Fan2015} leverage graph isomorphism properties to uniquely identify 
entities within a graph. Graph functional dependencies~\cite{Fan2016} apply attribute-value 
dependencies (similar to CFDs~\cite{Fan2008}) to the topological structures of graphs, while graph entity dependencies (GEDs)~\cite{Fan2019} further integrate the semantics of both GFDs and GKeys. 
In this study, we adopt graph differential dependencies (GDDs)~\cite{GDD}, an extension of GEDs. 
GDDs introduce similarity and matching semantics into the existing structure, making them 
suitable for defining declarative matching rules. This characteristic allows GDDs to better 
capture approximate matching relationships in graphs, thereby improving the accuracy and 
robustness of entity resolution (cf.~\cite{GDD, GraphER, LMMER}).


\noindent\underline{\em Progressive ER.} Madhavan et al.~\cite{madhavan2007web} were among the first
to implement a progressive data integration system in Google Base, demonstrating how web data 
can be integrated progressively under time and resource constraints. Since then, progressive 
methods have been widely applied to schema mapping~\cite{deuna2018machine, kimmig2017collective, 
kimmig2018collective, qian2012sample} and entity resolution~\cite{PSN,IPES}, especially in scenarios with limited computational capacity or time for debugging.
Existing progressive ER methods employ blocking and scheduling strategies to prioritize high-probability matches for faster discovery~\cite{IPES,PSN}. However, these approaches have two key limitations. First, since result emission is not performed at the entity level, matching remains incomplete, potentially leading to incorrect intermediate results that fail to meet user requirements. Second, it lacks support for graph data structures, e.g., entity profiles generated from partial cluster matches may result in inaccurate outcomes.

\noindent\underline{\em On-Demand ER.}
The most related works to this paper are in~\cite{QDA, BrewER, QueryER}. 
Query-Driven Approach (QDA)~\cite{QDA} streamlines data cleaning by analyzing selection 
predicates within a block to reduce comparisons. However, its lack of progressive execution 
capability makes it unsuitable for real-time applications. On the other hand,
BrewER~\cite{BrewER} incrementally cleans data during SQL SP query evaluation using prioritized 
attribute comparisons. While similar to our approach, it has three key limitations: it requires 
computationally expensive Cartesian products to generate candidates, demands whole-dataset 
blocking preparation, and lacks support for graph relationships.
QueryER~\cite{QueryER} addresses ER in the relational data setting through user-specified queries, operating within a well-defined schema using a dedicated parser, planner, and executor. Its goal is to integrate ER capabilities into query engines. In contrast, our work presents a standalone, on-demand ER solution for property graphs, without requiring integration into query processing systems.

\section{Preliminaries}
This section presents relevant concepts and definitions used in the paper. 

\subsection{Basic Definitions and Notions}
The definitions of {\em property graphs, graph patterns, and matches} follow~\cite{Fan2019}. 

A \textbf{\em property graph} is a directed graph \( G = (V, E, L, FA) \), where 
\( V \) is a set of nodes, \( E \subseteq V \times L \times V \) is a set of labeled edges, and each node \( v \in V \) has a unique ID, label \( L(v) \), and attributes \( FA(v) = [(A_1, c_1), \dots, (A_n, c_n)] \), where \( A_i \) are distinct attributes and \( c_i \) their values.

A \textbf{\em graph pattern} \( Q[\bar{z}] = (V_Q, E_Q, L_Q) \) describes node structures and relationships: \( V_Q \) is the set of pattern nodes, \( E_Q \) the set of edges, and \( L_Q \) a labeling function that assigns labels, including the wildcard \(*\), which denotes unrestricted labels. Two labels \( l_1 \) and \( l_2 \) match (\( l_1 \simeq l_2 \)) if \( l_1 = l_2 \) or either is a wildcard.

A \textbf{\em match} of a graph pattern $Q[\bar{z}]$ in a graph $G$ is a homomorphism, $h$,
mapping $Q$ onto the graph \( G\), preserving labels and structure.

An \textbf{\em entity profile} for a real-world entity is a tuple \( p = \langle pid, eid, type, P, R \rangle \), where \( pid \) is a unique profile identifier, \( eid \) is the (possibly unknown) real-world entity ID, and \( type \) denotes its category (e.g., \textit{person}, \textit{location}). \( P \) is a set of attribute-value pairs \( (A, c) \), and \( R \) a list of relations \( (rela, pid') \) linking the entity to other profiles. If two profiles, \(p_1, p_2\), represent the same entity, then \(p_1.eid = p_2.eid\).

\begin{figure}[!t]
    \centering
    \includegraphics[width=\linewidth]{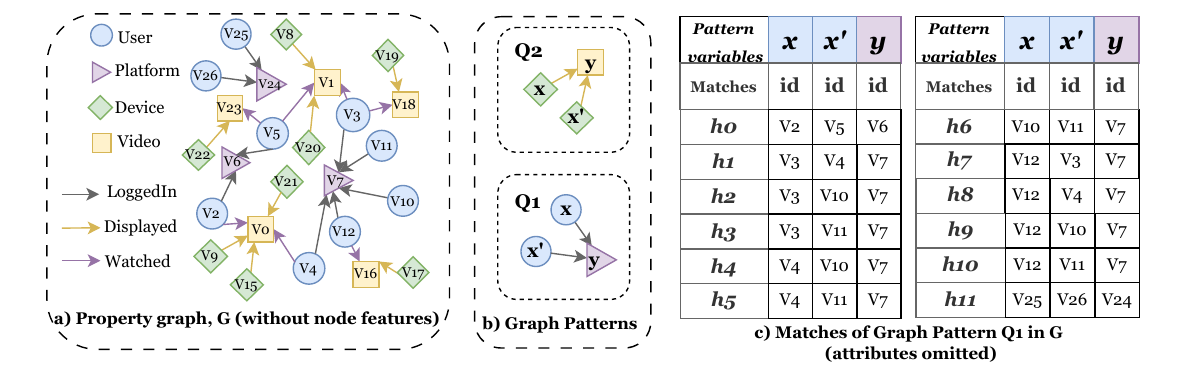}
    \caption{Toy Example: a) Property graph $G$; b) Graph patterns $Q_1, Q_2$; 
    c) Matches of $Q_1$ in $G$}
    \label{fig:toy}
\end{figure}

\begin{example}[Property graph, graph patterns, and matches]\label{ex:one}
Fig.~\ref{fig:toy} illustrates: a) a property graph, b) two graph patterns 
$Q_1[x,x',y]$ and $Q_2[x,x',y]$, and c) the matches of $Q_1$ in $G$ in a pseudo-table. 
As an example, the pattern $Q_1$ represents two \verb|user| nodes connected to the same \verb|platform| 
via a \verb|watched| relationship. Matches in $G$ are shown in Fig.~\ref{fig:toy}~c). 
\end{example}

\subsection{Graph Differential Dependency (GDD)}
We present the syntax and semantics of \textbf{\em graph differential dependency}~\cite{GDD}. 
A GDD \(\varphi\) is a pair \((Q[\bar{u}], \Phi_X \rightarrow \Phi_Y)\), where \(Q[\bar{u}]\) is a 
graph pattern, and \(\Phi_X, \Phi_Y\) are sets of distance constraints on pattern variables \(\bar{u}\). GDD captures dependencies among graph entities based on similarity or 
matching conditions.
The distance constraints are given as:
\begin{align*}
\delta_A(x.A, c) \leq t_A; \quad \delta_{A_1A_2}(x.A_1, x'.A_2) \leq t_{A_1A_2}; \quad 
\delta_{eid}(x.eid, c_e) = 0; \\
\quad \delta_{eid}(x.eid, x'.eid) = 0; \quad 
\delta_{\equiv}(x.rela, c_r) = 0; \quad \delta_{\equiv}(x.rela, x'.rela) = 0,
\end{align*}
where \(x, x' \in \bar{u}\) are pattern variables, \(A_i\) are attributes, and \(c\) is a constant. 
\(\delta_{A_1A_2}\) depends on attribute types (e.g., arithmetic operations for numerical values, edit distance for strings).
In GDD, \(\Phi_X\) specifies constraints that must hold for \(\Phi_Y\) to be satisfied. 
We use GDDs to encode user-specified and application-relevant constraints.
We illustrate GDD semantics with a toy example below. 

\begin{example}[GDD Semantics]\label{ex:gdd}
Consider $\varphi: (Q_1[x,x',y], \Phi_X\to \Phi_Y)$, where 
\(\Phi_X=\{\delta_{Age}(x,x')\leq 3 \wedge \delta_{Ph}(x,x')=0\}\) and 
\(\Phi_Y = \{\delta_{eid}(x,x')=0\}\). 
This states that if two \verb|users| in $Q_1$ have an \verb|Age| difference within 
\(3\) and identical \verb|Phone| numbers, they refer to the same real-world entity.
\qed
\end{example}

\subsection{Problem Definition}
Given a property graph, $G$, and a set $\Sigma$ of user-specified GDDs,
we seek to identify and link all $\Sigma$-relevant node pairs $v_i,v_j\in G$, 
that refer to the same real-world entity. A node pair $v_i,v_j\in G$
is said to be $\Sigma$-relevant iff: $v_i,v_j \in H(Q,G)\in G$, where
$H(Q,G)$ is the list of matches of a pattern $Q$ in $\Sigma$. 

The user-defined GDD specification is outside the scope of our study, but existing 
GDD mining algorithms~\cite{GDD,Zhang2023} can be used. 
We present empirical results in Section~\ref{sec:ab-rules} to guide rule selection.

\section{The FastER Framework for Property Graphs}
In this section, we introduce FastER, a framework designed for candidate filtering 
and match scheduling, allowing the integration of any matching function of your choice.
A brief pseudo-code is presented in Algorithm~\ref{alg:fastER}.
\begin{algorithm}[!t]
\captionsetup{belowskip=0pt}
\caption{{\sf FastER}($G, Q[\bar{z}], \delta, \text{threshold}$)}\label{alg:fastER}
\KwData{Property graph $G$}
\KwIn{Graph pattern $Q[\bar{z}]$,distance constraints $\delta$,edge weight threshold}
\KwResult{Matched results}
\tcc{Stage 1: Graph pattern filtering}
$\mathcal{H}_1 \gets \{\text{subgraphs} \mid \text{subgraphs satisfy } Q[\bar{z}]\}$ \\
\tcc{Stage 2: Constraint-based filtering}
$\mathcal{H}_2 \gets \{s \in \mathcal{H}_1 \mid s \text{ satisfies constraints } \delta\}$ \\
\tcc{Stage 3: Blocking graph construction and PPS}
\ForEach{$(x, x') \in \mathcal{H}_2$}{
    $edgeWeight \gets \sum_{r \in \mathcal{R}} \text{if } x, x' \text{ satisfy } r$ \\
    \If{$edgeWeight > 0$}{
        Add edge $(x, x')$ with $edgeWeight$ to $BlockingGraph$
    }
}
$SortedProfileList \gets \text{sortByEdgeWeights}(BlockingGraph)$ \\
\ForEach{$v \in SortedProfileList$}{
    $CompareList \gets \text{getCandidates}(v)$ \\
    \ForEach{$candidate \in CompareList$}{
        \If{$edgeWeight(v, candidate) \geq \text{threshold}$}{
            Compare($v, candidate$), updateMatchLists($v, candidate$)
            \\
            \If{meetsUserRequirements($v, candidate$)}{
    addToResultSet($v, candidate$)
}
        }
        \Else{
            prune($v, candidate$) \tcp*{Skip if edge weight is too low}
        }
    }
}
\Return{Final matched results}
\end{algorithm}
The FastER algorithm consists of three stages: graph pattern filtering, constraint-based 
filtering, and blocking graph construction and Progressive Profile Scheduling(PPS). In Stage 1, it identifies candidate 
subgraphs $\mathcal{H}_1$ that match the graph pattern $Q[\bar{z}]$ (line 1). In Stage 2, 
it refines these candidates by applying the user-defined constraints $\delta$, forming the 
set $\mathcal{H}_2$ (line 2). Finally, Stage 3 constructs a blocking graph from pairs in 
$\mathcal{H}_2$, computes edge weights based on rules satisfaction (lines 3-6), and sorts 
nodes to form a SortedProfileList (line 7). It iteratively processes nodes against candidate 
pairs, performing comparisons if their edge weight meets the specified threshold (lines 8-11), 
otherwise pruning the pair (line 16), the aggregated values at the entity level are validated 
to ensure they meet the user's demand (line 13). 
The algorithm returns the final matched results (line 17).

In addition, FastER offers interpretability by design. The filtering process is driven by user-defined GDD rules, which specify explicit graph patterns and attribute constraints. Each candidate match retained by the system can be traced back to the rules it satisfies, allowing users to understand and refine the matching behavior based on domain semantics.

\begin{figure}[!t]
  \centering
  \includegraphics[width=\textwidth]{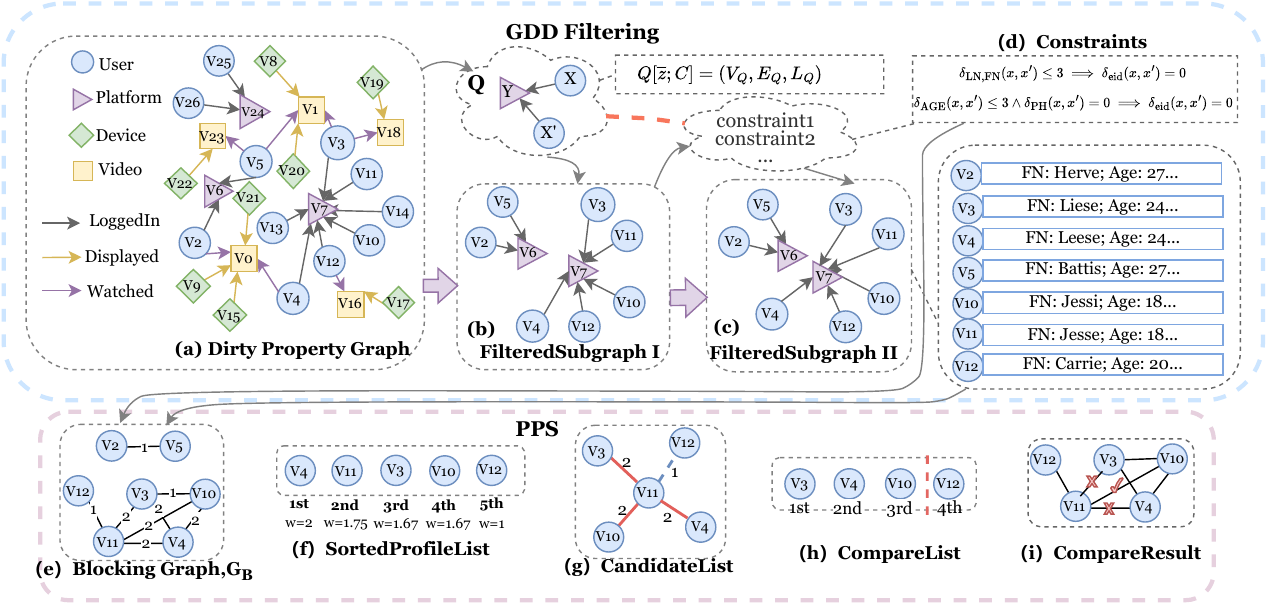}
  \caption{The overall architecture of FastER.
  (a) The dirty property graph to be cleaned, (b) Subgraph obtained after graph pattern filtering, (c) Subgraph further refined by constraint-based filtering, (d)Constraints in GDD, (e) Blocking graph generated based on candidate matches and constraints, (f) The sorted profile list, (g) The neighborhood of node V11 within the same block, (h) The comparison list after processing V11 during the emission phase, (i) Comparison results of V11's related candidates.}\label{fig:FastER}
\end{figure}

\subsection{Graph Pattern Filtering}
In the first stage of graph pattern filtering, we use graph patterns of the GDDs as query
inputs over the dirty property graph to extract subgraphs that satisfy the structural constraints. These graph patterns are designed to capture potential entity relationships based on the properties of nodes and the relationships between them.
For example, considering the property graph shown in Fig.~\ref{fig:FastER}~b), let us assume that we use the graph pattern defined in $Q[z; C] = (V_Q, E_Q, L_Q)$. The pattern specifies any two user nodes, $x$ and $x'$, connected to the same platform node $y$ via watch edges.
Distance constraints can be defined over this pattern. For example, nodes must satisfy certain attribute requirements, such as the user's age being greater than 18, which, under the definition of an on-demand query, ensures that only matches meeting specific user requirements are selected.
Based on this structure, we first extract a set of subgraph candidates from the original graph that satisfy this structural requirement. The set of matched subgraphs can be represented as:
\(
H_1(x, x', y) =
\{ (v_2, v_5, v_6), (v_3, v_4, v_7), (v_3, v_{10}, v_7), (v_3, v_{11}, v_7),
(v_4, v_{10}, v_7), (v_4, v_{11}, v_7), (v_{10},\\ v_{11}, v_7), 
(v_{12}, v_3, v_7), (v_{12}, v_4, v_7), (v_{12}, v_{10}, v_7), (v_{12}, v_{11}, v_7)\}.
\)

Compared to Fig.~\ref{fig:toy}, one fewer subgraph is present due to not satisfying the user's demand (i.e., on-demand).
These subgraphs, consisting of nodes and edges, satisfy the structure of user-platform relationships through ``logged-in'' edges,and the user-specified demand, forming the candidate subgraphs for further filtering in the next step.

\subsection{Constraints-Based Filtering}
Here, we apply a predefined set of distance constraints to further filter 
matches of the subgraphs from the first stage. These constraints, based on domain knowledge, are designed to ensure that candidate node pairs exhibit sufficient similarity, thereby retaining only the most promising matching candidates. For example, the distance constraints may require that the last names or phone numbers of two nodes be similar for them to be considered potential matches.

We define several distance constraints, which require that the differences between node pairs on certain attributes must fall within specific thresholds, such as last names, phone numbers, and ages.
In this example, we define four constraints:  
- Lastname (LN) edit distance: $\delta_{\text{LN}}(x, x') \leq 3$  
- Firstname (FN) edit distance: $\delta_{\text{FN}}(x, x') \leq 3$  
- Phone (PH) edit distance: $\delta_{\text{PH}}(x, x') \leq 3$  
- Age (AGE) difference: $\delta_{\text{AGE}}(x, x') = 0$. Only candidate pairs that satisfy at least one of these constraints will be retained for further processing. 

After applying these filtering constraints, we obtain a more refined set of candidate pairs and their corresponding subgraphs. These subgraphs focus on nodes and edges that satisfy the similarity constraints and are therefore more likely to represent entity matches.


\subsection{Blocking Graph Construction and PPS} 
Next, we construct a Blocking Graph based on the refined candidate list. In this graph, nodes represent entities, and edges represent potential matches based on their attributes. The weight of each edge is determined by the similarity between connected nodes. For simplicity and clarity, we assume that if two nodes satisfy one rule, the edge weight increases by one. We can define a Blocking Graph as shown in Fig.~\ref{fig:FastER}~e). This graph naturally divides the nodes from the original subgraph into two blocks, reducing the number of node comparisons and improving matching efficiency.

However, to improve matching accuracy, more sophisticated weighting methods can be applied. For example,  the ARCS function is used to assign weights by summing the inverse cardinality of common blocks. The ARCS function assigns higher weights to pairs of entities that share smaller (i.e., more distinctive) blocks, and is defined as:
\(
ARCS(p_i, p_j, B) = \sum_{b_k \in B_i \cap B_j} \frac{r_{p_i, p_j, b_k}}{|b_k|},
\)
where \(r_{p_i, p_j, b_k}\) denotes the number of rules satisfied by the entity pair \( (p_i, p_j) \) within the common block \( b_k \). This modification ensures that pairs satisfying more rules receive proportionally higher weights while still considering the distinctiveness of smaller blocks. Similarly, other weighting functions can be applied to the Blocking Graph to assign higher weights to edges connecting nodes with strong co-occurrence patterns and lower weights to edges representing casual co-occurrences. 
Thus, with this weighting mechanism, the Blocking Graph not only reflects the similarity between entities more accurately, but it also adjusts the matching priorities based on the importance of different attributes, making the matching process more efficient overall.

\noindent\underline{\em Progressive Profile Scheduling (PPS).}
Once the Blocking Graph is constructed, we adopt the Progressive Profile Scheduling (PPS) technique to efficiently perform entity matching. PPS is an entity-centric scheduling method designed to prioritize nodes with the highest likelihood of matching by assessing their duplication probability.

First, within each block, nodes are sorted based on the average weight of their edges in the Blocking Graph, generating a SortedProfileList. The nodes in this list are ranked from highest to lowest, and nodes with higher average edge weights are prioritized for comparison, as they are more likely to match with other nodes.
For instance, in Fig.~\ref{fig:FastER}~c), each node has multiple candidate nodes. We begin by comparing the node with the highest weight in the sorted list. During this process, PPS prunes unnecessary comparisons based on a predefined threshold (e.g., edge weight less than 2). For example, in the case of node \(v_{11}\), if the edge weight between node \(v_{12}\) and \(v_{11}\) is below the threshold, this comparison is skipped, and \(v_{12}\) is removed from \(v_{11}\)'s candidate list.

By utilizing this selective pruning strategy, PPS significantly reduces the number of required comparisons, thereby improving matching efficiency. In this example, edges with a weight of 1 are ignored, leading to the removal of \( 
v_2 \leftrightarrow v_5 \text{ (weight 1)}, \quad
v_{12} \leftrightarrow v_{11} \text{ (weight 1)}, \quad
v_3 \leftrightarrow v_{10} \text{ (weight 1)}
\).
The remaining comparisons are: \( 
v_3 \leftrightarrow v_{11} \text{ (weight 2)}, \quad
v_4 \leftrightarrow v_{11} \text{ (weight 2)}, \quad 
v_4 \leftrightarrow v_{10} \\ \text{ (weight 2)},\quad
v_{11} \leftrightarrow v_{10} \text{ (weight 2)}, \quad
v_3 \leftrightarrow v_4 \text{ (weight 2)}
\). Consequently, the total number of comparisons is reduced from 8 to 5, as the three edges with a weight of 1 are ignored. Before publishing the final results, the aggregated values at the entity level will be validated to ensure they meet the user's demand.

\noindent\underline{\em Transitive Matching Optimization.} In an ideal scenario, transitive relationships can be leveraged to further reduce the number of comparisons. For example, if both \(v_3\) and \(v_4\) match \(v_{11}\), and \(v_{11}\) also matches \(v_{10}\), we can infer through transitivity that \(v_3\), \(v_4\), \(v_{10}\), and \(v_{11}\) all represent the same entity. As a result, redundant comparisons can be eliminated, leaving only the essential matches: \( 
v_4 \leftrightarrow v_3, \quad v_4 \leftrightarrow v_{10}, \quad v_3 \leftrightarrow v_{11}
\). This reduces the total number of comparisons from 5 to 3, thereby optimizing the resolution process.
\subsection{Time Complexity}

Let \(G = (V, E)\) be a property graph with \(|V| = n\) nodes and \(|E| = m\) edges, and let \(Q = (V_Q, E_Q)\) be a small graph pattern. Subgraph isomorphism using Cypher (the query language used by Neo4j) in Neo4j has worst-case time complexity \(O(m \cdot |E_Q|)\), which reduces to \(O(m)\) for constant-size patterns (as discussed in Appendix A.1, Section 1).

Let \(C\) be the number of candidate node pairs identified by pattern matching. Each of these undergoes \(d\) GDD constraint checks, giving \(O(C \cdot d)\) time. Since \(d\) is typically constant, this simplifies to \(O(C)\). After constraint filtering, only a fraction \(\beta \in (0, 1)\) of the pairs remain. Let \(C' = \beta C\) denote the number of valid pairs after filtering.
Blocking graph construction on these \(C'\) pairs requires \(O(C' \cdot r)\) for edge weight computation, where \(r\) is the number of features per edge, and \(O(C' \log C')\) for sorting by similarity score.
Finally, Progressive Profile Scheduling (PPS) incrementally compares each query-relevant profile (total \(N\)) to its \(k'\) most promising candidates (typically \(k' \ll C'\)), yielding \(O(N \cdot k')\) complexity for the matching phase.

Summing the steps, the total complexity is:
\(
O(m + \beta C + C' \log C' + N \cdot k').
\)
As shown in Appendix A.1, \(\beta\) and \(k'\) are empirically small in most cases (e.g., \(\beta < 0.1\), \(k' < 5\)), yielding an overall complexity close to linear in the graph size \(m\) and number of target entities \(N\).

\section{Experimental Evaluation}
This section empirically evaluates our proposal {\em w.r.t.}
different SOTA solutions. 

\subsection{Experimental Setting} 
All experiments were performed on a PC with a 2.40GHz Intel Xeon Silver 4210R 
processor and 32GB of memory, running Linux OS. The experiments were implemented in 
Python 3.10 and interacted with the Neo4j graph database (version 5.19.0).

\noindent\underline{Datasets.} In this study, we used five graph benchmark datasets and five relational benchmark datasets (cf.~Table~\ref{tab:combined}) from various domains.  
All datasets used in this paper are publicly available on the GitHub repository\footnote{\url{https://github.com/Zaiwen/On_Demand_Entity_Resolution  }}. 
Although FastER is specifically designed for property graphs, we show it can also be applied in the relational data setting. Thus, we transformed the relational datasets in Table~\ref{tab:combined} into graph representations by leveraging their attribute relationships. The detailed conversion methods are provided in the ~\href{https://github.com/Zaiwen/On_Demand_Entity_Resolution}{GitHub repository}. The reported values for Nodes, Edges, Node Types, and Edge Types in the relational datasets are based on their graph-transformed statistics.
We apply automated GDD mining algorithms~\cite{GDD,Zhang2023}, followed by manual selection of a small number of high-quality rules to ensure interpretability and relevance. This process is detailed in subsection~\ref{sec:ab-rules}. To ensure a fair and consistent comparison across methods, all approaches use the same oracle matcher.

\noindent\underline{Methods.} We categorize the examined methods into On-Demand ER, Progressive ER, and Batch ER methods. \textbf{On-Demand ER methods} include BrewER and QDA. BrewER~\cite{BrewER} is an on-demand framework that enables direct execution of SQL SP queries on dirty data while progressively returning query results. QDA~\cite{QDA} is a query-driven algorithm that optimizes matches within a data block by using selective predicates to reduce unnecessary record comparisons. \textbf{Batch ER methods} include HG, RobEM, and Ditto.
HG~\cite{HG} employs a hierarchical graph attention transformation network for ER tasks. RobEM~\cite{Rastaghi2022} leverages pre-trained language models (PLMs) for entity resolution, while Ditto~\cite{Ditto} is a Transformer-based PLM approach designed for ER. \textbf{Progressive ER methods} include PSN and I-PES. PSN~\cite{PSN} employs dynamic window adjustment to prioritize and resolve matching entities by initially sorting and selecting the most relevant records. I-PES~\cite{IPES} is an incremental ER algorithm that dynamically maintains the comparison order using a priority queue to achieve progressive resolution. All hyperparameters are set to the default values from the original papers unless otherwise specified.

\noindent\underline{Evaluation Metrics.}
It is important to note that in our experiments, we do not directly evaluate precision or F1-score. 
FastER is a matcher-agnostic framework that focuses on filtering and scheduling components. 
Therefore, we adopt an oracle matcher—i.e., entity pairs are labeled using ground truth—
to isolate the effects of the matching function and enable a fair comparison of the framework’s 
efficiency and recall capabilities.
Additionally, following standard practice in ER on-demand literature~\cite{BrewER}, the final matching 
confirmation stage guarantees a precision of 1. As a result, we exclude precision and F1-score from our 
evaluation. Instead, we focus on metrics such as query recall and error rate, which are discussed in detail later.
We explore the following questions in subsequent subsections:
\begin{enumerate}
    \item How does FastER perform in ER-on-demand scenario (Exp-1)?
    \item How does FastER compare to  SOTA batch processing methods (Exp-2)?
    \item How well do progressive solutions perform in on-demand ER tasks? (Exp-3)
    \item How does FastER perform on different components (Exp-4)?
    \item How can users be guided to select appropriate rules (Exp-5)?
\end{enumerate}

\begin{table}[!t]
    \centering
    \caption{Graph and Relational Benchmark Datasets (NT: Node Types, ET: Edge Types, ND: Node Duplicates, ER: Entity Resolution}
    \label{tab:combined}
    \arrayrulecolor{gray}
    \begin{tabular}{c|c|c|c|c|c|c|c}
        \hline
        Category & Dataset & Nodes & Edges & NT & ET & ND & Domain \\
        \hline
        \specialrule{0.08em}{0pt}{0pt} 
        Graph & WWC & 2.688k & 15.757k & 5 & 9 & 202 & Sports \\
        \cline{2-8}
        & GDS & 8.977k & 80.365k & 5 & 5 & 350 & Airport \\
        \cline{2-8}
        & Entity Resolution (ER) & 1.237k & 1.819k & 4 & 3 & 12 & Streaming \\
        \cline{2-8}
        & ArXiv & 88.07k & 58.515k & 2 & 1 & 5.92k & Research \\
        \cline{2-8}
        & CiteSeer & 4.393k & 2.892k & 2 & 1 & 456 & Citations \\
        \hline
       Relational & SIGMOD20 & 13.58k & 12.01k & 3 & 2 & 3.06k & Cameras \\
        \cline{2-8}
        & Altosight & 12.47k & 12.44k & 3 & 2 & 453 & Cameras \\
        \cline{2-8}
        & Fodors-Zagats (FZ) & 1.69k & 1.73k & 3 & 2 & 112 & Restaurants \\
        \cline{2-8}
        & DBLP-ACM (DA) & 8.70k & 9.82k & 3 & 2 & 2.22k & Bibliography \\
        \cline{2-8}
        & Amazon-Google (AG) & 4.98k & 4.59k & 2 & 1 & 116 & Software \\
        \hline
    \end{tabular}
    \arrayrulecolor{black} 
\end{table}

\subsection{Exp-1: FastER vs On-Demand Baselines}
Since FastER is a matcher-agnostic framework, we adopt an oracle matcher across all three methods to better evaluate its performance. This matcher accurately labels all comparisons based on the known ground truth.
As both BrewER 
and QDA are frameworks that can embed any blocking function, we employ JedAI~\cite{Papadakis2020,Papadakis2019}, which is based on a completely unsupervised blocking approach. We use its standard 
configuration, which relies on Token Blocking and Meta-blocking~\cite{Papadakis2020}. 
This blocking method has been proven to achieve good results with BrewER and QDA.

The reported performance of BrewER in our experiments may differ from that in its original paper~\cite{BrewER}, as the original study relies on ground truth for blocking, whereas we employ the unsupervised approach JedAI to better reflect real-world scenarios.
 This setup ensures a fair comparison with FastER.

To comprehensively evaluate performance, we executed two queries on different datasets, each with 
distinct attribute constraints, and generated ground truth matching sets for each subgraph.  
To assess effectiveness, we introduce \textit{query recall}, which considers only the subset of interest 
to the user, rather than the entire result set. It is defined as:
\(
R_{\text{query}} = \frac{|M_{\text{query}} \cap M_{\epsilon}|}{|M_{\text{query}}|}
\)
where $M_{\text{query}}$ is the ground truth matching set within the subset, and $M_{\epsilon}$ 
is the set of matches found by the algorithm in the same subset. 

\begin{figure*}[!t]
    \centering
    \includegraphics[width=\textwidth]{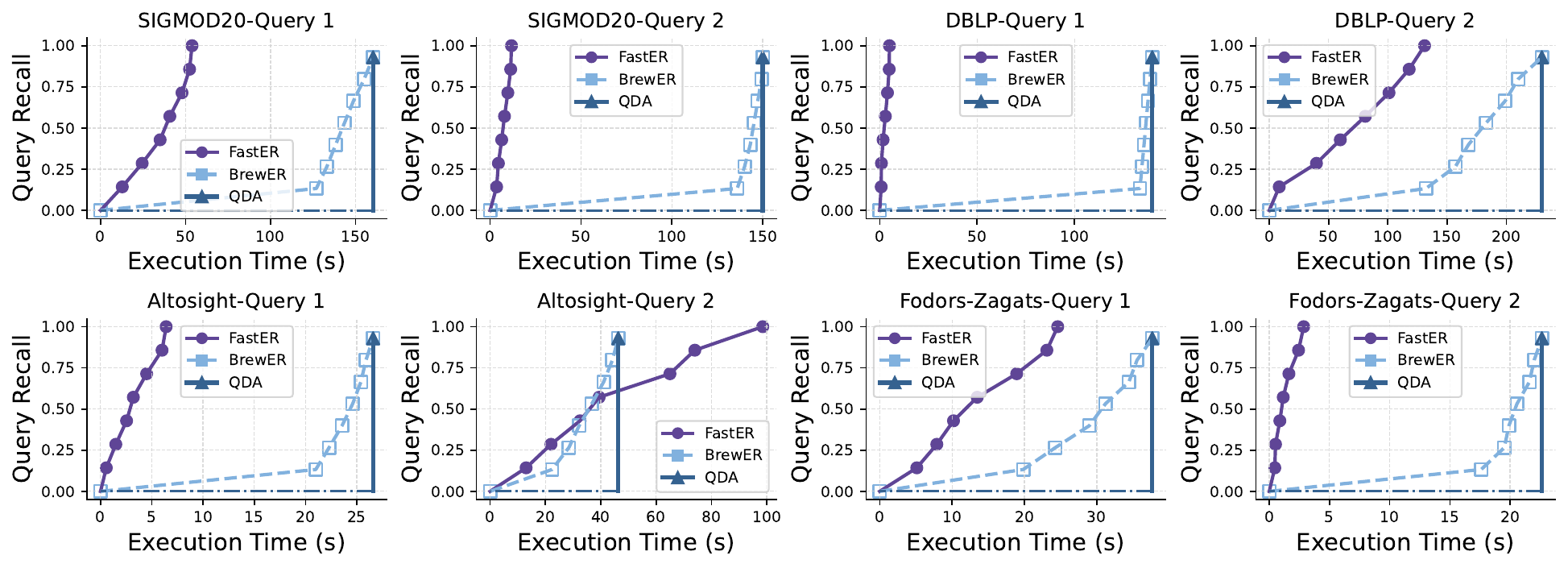}
    \caption{FastER vs On-Demand baselines}
    \label{fig:fo}
\end{figure*}

\begin{wrapfigure}[10]{r}{0.5\textwidth}
\vspace{-26pt}
    \centering
    \includegraphics[width=0.5\textwidth]{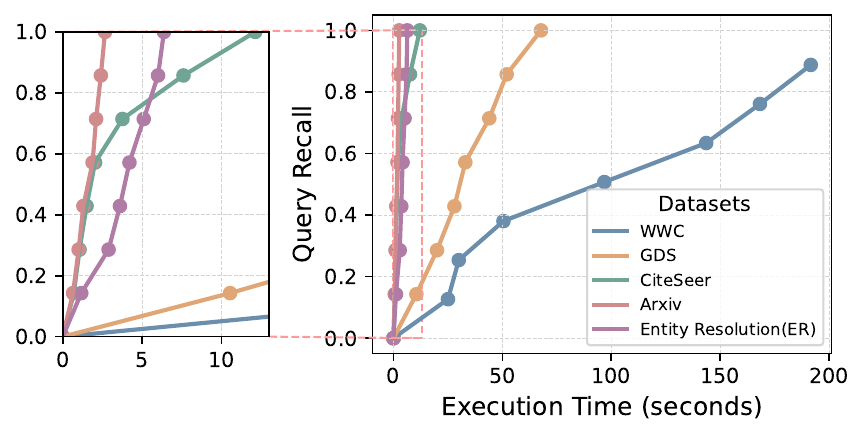}
    \caption{Performance of FastER on Graphs}
    \label{fig:grr}
\end{wrapfigure}
First, we show how FastER performs on all graph data sets in Table~\ref{tab:combined} -- 
captured in Fig.~\ref{fig:grr}. The results highlight the real-time capability of
FastER. In general, graph complexity dictates efficiency as the characteristics in
Table~\ref{tab:combined} bear out.
Further, Fig.~\ref{fig:fo} shows the relative performance of FastER \emph{w.r.t.} on-demand
baselines. The plots show QDA exhibits a typical stepwise execution curve because it must compare all candidate pairs before producing any output. In contrast, the recall rates of both FastER and BrewER show a gradual increase across all datasets.

Additionally, FastER generally achieves a lower execution time by performing blocking after filtering, reducing blocking overhead. In contrast, BrewER partitions the entire dataset, incurring higher preparation costs.
FastER underperforms on \textit{Altosight-Query2} because the rules fail to effectively capture candidates, leading to insufficient filtering.

\begin{table}[!t]
\centering
\small
\caption{FastER vs Batch-Query baseline}
\label{tab:combined_comparison}
\arrayrulecolor{gray}

\resizebox{\linewidth}{!}{
\begin{tabular}{l|cccc|cccc}
    \toprule
    Dataset & 
    \multicolumn{4}{c|}{Recall} & 
    \multicolumn{4}{c}{Tavg (k seconds)} \\
    \cmidrule(lr){2-5} \cmidrule(lr){6-9}
    & HG & RobEM & Ditto & FastER & HG & RobEM & Ditto & FastER \\
    \midrule
    FZ 
    & 0.954 & \textbf{1.000} & \textbf{1.000} & \textbf{1.000} 
    & $2.781\,\scalebox{0.75}{$\pm\,0.053$}$ & $4.349\,\scalebox{0.75}{$\pm\,0.076$}$ & $0.917\,\scalebox{0.75}{$\pm\,0.015$}$ & $\mathbf{0.553\,\scalebox{0.75}{$\pm\,0.011$}}$ \\

    DA
    & 0.982 & 0.982 & 0.978 & \textbf{1.000} 
    & $1.440\,\scalebox{0.75}{$\pm\,0.042$}$ & $3.198\,\scalebox{0.75}{$\pm\,0.088$}$ & $0.464\,\scalebox{0.75}{$\pm\,0.009$}$ & $\mathbf{0.052\,\scalebox{0.75}{$\pm\,0.003$}}$ \\

    AG
    & 0.654 & \textbf{0.846} & 0.786 & 0.841 
    & $1.162\,\scalebox{0.75}{$\pm\,0.033$}$ & $5.245\,\scalebox{0.75}{$\pm\,0.121$}$ & $\mathbf{0.818\,\scalebox{0.75}{$\pm\,0.018$}}$ & $2.910\,\scalebox{0.75}{$\pm\,0.077$}$ \\

    WWC
    & 0.902 & 0.585 & \textbf{0.927} & 0.923
    & $1.069\,\scalebox{0.75}{$\pm\,0.029$}$ & $4.270\,\scalebox{0.75}{$\pm\,0.089$}$ & $1.105\,\scalebox{0.75}{$\pm\,0.022$}$ & $\mathbf{0.630\,\scalebox{0.75}{$\pm\,0.013$}}$ \\

    GDS 
    & 0.943 & 0.800 & 0.900 & \textbf{1.000} 
    & $1.817\,\scalebox{0.75}{$\pm\,0.040$}$ & $4.665\,\scalebox{0.75}{$\pm\,0.102$}$ & $0.678\,\scalebox{0.75}{$\pm\,0.014$}$ & $\mathbf{0.445\,\scalebox{0.75}{$\pm\,0.010$}}$ \\
     
    \bottomrule
\end{tabular}
}
\end{table}
\subsection{Exp-2: FastER vs Batch-Query Baseline}
To evaluate FastER against batch processing methods, we define "average time per matching result" (Tavg) as a fair metric for speed comparison, and we still extract a sub-ground-truth. Tavg is computed as:
\(
T_{\text{avg}} = \frac{T_{\text{total}}}{\left| M_{\text{emitted}} \cap M_{\text{gt}} \right|},
\)
where \(T_{\text{total}}\) is the total query execution time, 
\(M_{\text{emitted}}\) is the set of entity matches that have been emitted, 
and \(M_{\text{gt}}\) represents the sub-ground-truth matching set.

Table~\ref{tab:combined_comparison} compares FastER and batch-query baselines across multiple datasets, focusing on recall and average time per matching result (Tavg). FastER consistently achieves the highest recall, demonstrating the effectiveness of rule filtering in capturing matching candidates. Regarding time efficiency, batch-query baselines require full dataset traversal, whereas our approach significantly enhances practical performance. Tavg comparisons show FastER achieves substantial speed improvements across most datasets, except for Amazon-Google. The reason for this exception lies in the large intra-cluster variance of the dataset, which makes it difficult for the rules to accurately capture all matching candidates. Attempting to capture all candidates would come at the cost of introducing a large number of irrelevant matches. Therefore, users need to carefully balance and weigh their decisions when mining and selecting rules. Specific optimization suggestions will be discussed in detail in subsection~\ref{sec:ab-rules}.

\begin{figure*}[!t]
    \centering
    \includegraphics[width=\textwidth]{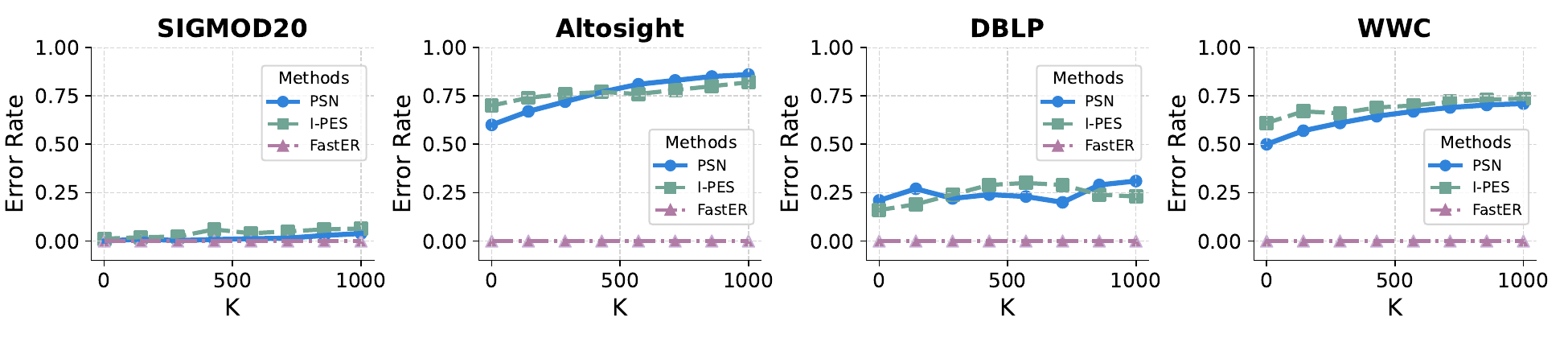}
    \caption{FastER vs Progressive-Query baseline (error rate)}
    \label{fig:fp}
\end{figure*}

\begin{table}[!t]
\centering
\caption{Ablation study results on graph and relational datasets.}
\label{tab:fastER_ablation}
\scriptsize
\resizebox{\linewidth}{!}{
\begin{tabular}{
    l
    | S[table-format=6.1] S[table-format=3.2]
    | S[table-format=5.1] S[table-format=3.2]
    | S[table-format=3.1] S[table-format=3.2]
}
\toprule
\multicolumn{7}{c}{\textbf{Graph Datasets}} \\
\toprule
Method & 
\multicolumn{2}{c|}{Arxiv} & 
\multicolumn{2}{c|}{CiteSeer} & 
\multicolumn{2}{c}{WWC} \\
\cmidrule(lr){2-3} \cmidrule(lr){4-5} \cmidrule(lr){6-7}
& {Rel. Comp.} & {Recall~(\%)} 
& {Rel. Comp.} & {Recall~(\%)} 
& {Rel. Comp.} & {Recall~(\%)} \\
\midrule
No-RF   & 453508.8 & 99.67 & 14659.1 & 99.20 & 117.9 & 99.00 \\
No-B    & 10.4     & 98.96 & 29.4    & 98.67 & 3.0   & 89.90 \\
No-PPS  & 1.6      & 97.78 & 1.4     & 97.09 & 2.3   & 89.20 \\
No-T    & 2.9      & 97.68 & 3.0     & 96.42 & 2.2   & 89.20 \\
FastER  & 1.0      & 97.68 & 1.0     & 96.42 & 1.0   & 89.20 \\
\bottomrule
\end{tabular}
}
\vspace{1em}
\resizebox{\linewidth}{!}{
\begin{tabular}{
    l
    | S[table-format=4.1] S[table-format=3.2]
    | S[table-format=5.1] S[table-format=3.2]
    | S[table-format=3.1] S[table-format=3.2]
}
\toprule
\multicolumn{7}{c}{\textbf{Relational Datasets}} \\
\toprule
Method & 
\multicolumn{2}{c|}{Fodors-Zagat} & 
\multicolumn{2}{c|}{DBLP-ACM} & 
\multicolumn{2}{c}{Amazon-Google} \\
\cmidrule(lr){2-3} \cmidrule(lr){4-5} \cmidrule(lr){6-7}
& {Rel. Comp.} & {Recall~(\%)} 
& {Rel. Comp.} & {Recall~(\%)} 
& {Rel. Comp.} & {Recall~(\%)} \\
\midrule
No-RF   & 1210.2 & 100.00 & 11662.8 & 100.00 & 301.1 & 99.00 \\
No-B    & 8.2    & 100.00 & 3.1     & 100.00 & 6.7   & 87.00 \\
No-PPS  & 1.2    & 100.00 & 1.6     & 100.00 & 1.4   & 82.18 \\
No-T    & 2.2    & 100.00 & 1.5     & 100.00 & 1.8   & 79.00 \\
FastER  & 1.0    & 100.00 & 1.0     & 100.00 & 1.0   & 79.00 \\
\bottomrule
\end{tabular}
}
\end{table}

\subsection{Exp-3: FastER vs Progressive-Query Baselines}
The most advanced progressive ER methods, such as \textit{PSN}~\cite{PSN} and \textit{I-PES}~\cite{IPES}, rank entity pairs by matching likelihood, prioritizing the most probable matches to enhance early quality. However, new matches may alter the aggregated results, leading to inconsistencies. Unlike FastER, progressive baseline lacks a final verification mechanism to ensure strict on-demand output, contributing to its high error rate. We evaluate progressive baseline's accuracy using the error rate over the first \(k\) emitted entities, denoted as \(\text{Err@}k\):
\(
\text{Err@}k = \frac{\text{Errors}}{k}.
\)
Here, an \textit{error} refers to an emitted result that does not meet user requirements, quantifying the proportion of incorrect matches in the first \(k\) outputs.

Fig.~\ref{fig:fp} shows error rate variations across datasets and \(k\) values. FastER, as a precise method, consistently maintains a \mbox{0\%} error rate by verifying matches individually. In contrast, progressive baseline exhibits high error rates in most datasets, except for SIGMOD20, which has minimal intra-cluster variance. Thus, progressive baseline is unreliable for ER-on-demand.

\subsection{Exp-4: Ablation Study}
In this experiment, we conducted an ablation study on the FastER framework to evaluate the impact of different components on matching efficiency and recall across multiple graph and relational datasets. 
We systematically removed the key components as follows.
1) No Rules Filtering (No-RF): Since blocking relies on rules filtering, this renders the system equivalent to a pure PPS method without blocking. 
2) No Blocking (No-B): Blocking partitions the matching space to minimize unnecessary comparisons. 
3) No Progressive Profile Scheduling (No-PPS): PPS enables the system to incrementally generate matches and employs pruning strategies to reduce the number of comparisons. 
4) No Transitivity Matching (No-T): Transitivity matching leverages existing matches to infer new ones, enhancing  completeness while reducing the number of comparisons.

The results are presented in Table~\ref{tab:fastER_ablation} showing the relative number of comparisons in each
ensuing method with respect to FastER's number of comparison.
In general, the results indicate that rules filtering is crucial for reducing the number of comparisons, as its removal significantly increases computational overhead. On datasets with low variance, eliminating rules filtering has minimal impact on recall, whereas on high-variance datasets, the effect is relatively significant. Blocking significantly affects execution time; removing it leads to a substantial increase in the number of comparisons, but has little impact on recall. PPS plays a key role in system responsiveness; its pruning mechanism effectively reduces the number of queries. Without PPS, processing time increases, though recall remains largely unaffected. Transitivity matching effectively reduces the number of comparisons without compromising recall.

\begin{figure*}[!t]
    \centering
    \includegraphics[width=\textwidth]{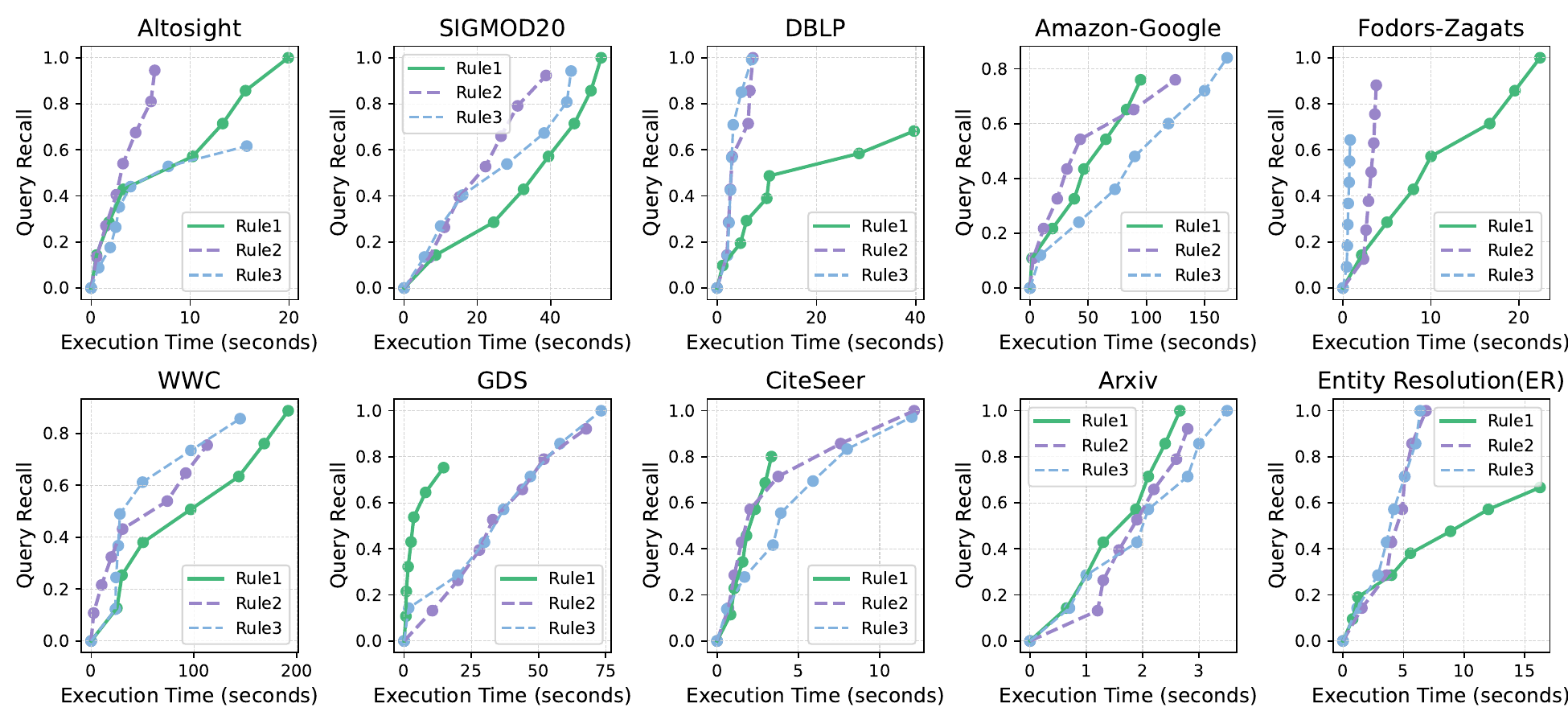}
    \caption{Ablation study of the impact of multiple rules on filtering performance}
    \label{fig:multirules}
\end{figure*}

\subsection{Exp-5: Rules Selection Strategies for Efficient Entity Resolution}
\label{sec:ab-rules} 
We conducted experiments on ten datasets, applying three different filtering rules to each. The specific rules used are provided in Appendix A.2.

The results (Fig.~\ref{fig:multirules}) show that rules with steeper slopes and higher recall are of higher quality, as they achieve superior performance within shorter execution times. Based on our findings, we summarize key principles for selecting filtering rules.
First, the strictness of rules impacts execution time and recall. Looser rules increase candidate matches, leading to longer execution times, while overly strict rules reduce execution time but risk excessive filtering, lowering recall. Thus, selecting appropriate rules is essential to balance efficiency and accuracy.

Moreover, as the complexity of relationships among nodes increases, the graph pattern should incorporate more nodes, and filtering rules should constrain multiple attributes to effectively capture structural dependencies.

When selecting a graph pattern, overly strict constraints should be avoided. For example:
\textit{Pattern 1}: \(x \rightarrow y \rightarrow z \leftarrow y' \leftarrow x'\);
\textit{Pattern 2}: \(x \rightarrow y \rightarrow z\), \(x' \rightarrow y' \rightarrow z'\).
Requiring node \(z'\) to be strictly identical to \(z\) may exclude many valid candidates. Instead, constraining the similarity between \(z\) and \(z'\) rather than enforcing identical entity IDs (eid) prevents over-filtering and enhances match coverage.
In summary, filtering rule and graph pattern selection directly influence the trade-off between execution time and recall. Stricter rules are preferable when execution time is a priority, whereas relatively looser rules are recommended when balancing execution time and recall. FastER is designed to leverage domain-specific rules for high-precision filtering. 
Meanwhile, recent work has begun to explore automated strategies for rule selection in graph-based entity resolution. 
We provide supporting resources on our project page\footnote{\url{https://github.com/Zaiwen/On_Demand_Entity_Resolution}}, 
which assist users in automatically constructing effective rules programmatically. This line of research is actively evolving and offers potential for future enhancements.

\section{Conclusion}
We introduce FastER, an efficient on-demand entity resolution (ER) approach for property graphs based on Graph Differential Dependencies (GDDs), addressing the six key challenges outlined in Figure 1. FastER captures complex graph patterns for high-precision entity matching and employs Progressive Profile Scheduling (PPS) to incrementally output results and provide real-time feedback.  
To enhance computational efficiency, FastER applies rule-based filtering, reducing candidate entity pairs by a factor of \(10^3\) to \(10^5\), and prioritizes user queries, focusing on user-specified subgraphs to avoid unnecessary computations. Experimental results demonstrate that FastER significantly outperforms existing ER methods in both processing time and recall.  
Future research could explore advanced filtering techniques for heterogeneous data sources and automated rule selection mechanisms to better handle intra-cluster variations.

\subsubsection*{Acknowledgment.}
The authors affiliated with institutions \(1, 5\) and \(6\) were supported in part by the National Key Research and Development Program of China under Grant No. 2023YFF1000100; the Hubei Key Research and Development Program of China under Grant Nos. 2024BBB055 and 2024BAA008; the Major Science and Technology Project of Yunnan Province under Grant No. 202502AE090003; and the Fundamental Research Funds for the Central Universities under Grant No. 2662025XXPY005.

\subsubsection*{Supplemental Material Statement.} Our supplementary material is available at \url{https://doi.org/10.5281/zenodo.15389843} and includes the source code and datasets used for running FastER. All details are provided in the \texttt{README.md} file. \texttt{Appendix.pdf} contains the algorithm complexity proof and Cypher queries.

\appendix

\input{appendix.tex}

\end{document}

%% file: appendix.tex
\section*{Appendix A.1: Time Complexity Analysis}
\noindent \textbf{(1) Neo4j-Based Subgraph Isomorphism Computation:} Subgraph isomorphism in a property graph $G$ with $n$ nodes and $m$ edges, given a query pattern $Q$ with $|V_Q|$ nodes and $|E_Q|$ edges, is in general an NP-hard problem. However, when using Neo4j (via Cypher queries) for pattern matching, the complexity can be approximated by scanning the graph for each element of the pattern. In the worst case, the query engine may have to traverse $O(m)$ relationships for each of the $|E_Q|$ edges in $Q$, leading to a time complexity on the order of $O(m \cdot |E_Q|)$. In practice, $|E_Q|$ is typically much smaller than $m$ (the patterns derived from Graph Differential Dependencies are small), so this cost approaches $O(m)$ – essentially linear in the size of the graph. (If the query pattern includes labels or indexed properties, Neo4j can leverage them to narrow down matches, but in the absence of highly selective indices, one can assume a linear scan per pattern edge in the worst case.) Thus, executing the subgraph isomorphism query in Neo4j will generally scale linearly with the graph size for small patterns, but it could grow linearly with $|E_Q|$ for more complex patterns. 

\noindent \textbf{(2) Impact of GDD Filtering and Blocking:} FastER introduces Graph Differential Dependencies (GDDs) as constraints to drastically reduce the search space before performing expensive comparisons. Instead of searching the entire graph $G$, GDDs focus the computation on a query-relevant subgraph. Concretely, suppose applying the GDD rules yields a set of $C$ candidate node pairs (potentially matching entity pairs), where $C \ll m$. This means only $C$ pairs need to be considered further, as opposed to $\Theta(n^2)$ or other large combinations without filtering. Checking each candidate pair against the set of $d$ GDD constraints incurs a cost of $O(C \cdot d)$ in total (each constraint check is $O(1)$ or a small constant factor, so this is effectively $O(C)$). After evaluating all GDD conditions, only a fraction of the pairs survive the filtering. Let $C'$ denote the number of pairs remaining after GDD filtering; empirically we can express $C' = \beta\,C$ with a small fraction $\beta \ll 1$. These $C'$ filtered candidates are then organized using a blocking strategy. FastER builds a blocking graph where nodes represent entity profiles and edges represent candidate matches between profiles. Constructing this graph and computing a similarity weight for each of the $C'$ edges takes $O(C' \cdot r)$ time, where $r$ is the number of attributes or features used to calculate the edge weight (this is linear in $C'$ if $r$ is a fixed small number of features). After weight computation, the candidate pairs (edges) are sorted by their similarity score, which adds an $O(C' \log C')$ overhead. The combination of GDD-based filtering and blocking ensures that the number of comparisons to be performed is vastly reduced and that those comparisons are structured: instead of considering all possible pairs, we only deal with $C'$ high-likelihood pairs, and we have them sorted by likelihood for efficient processing. In summary, these steps reduce the problem size from $m$ (or worse) down to $C'$ and impose only linear or near-linear overheads in doing so. 

\noindent \textbf{(3) Progressive Profile Scheduling (PPS) Optimization:} After filtering and blocking, FastER employs Progressive Profile Scheduling (PPS) to perform the actual entity matching in an efficient, incremental fashion. PPS processes the sorted list of candidate pairs in descending order of similarity, which means it attempts the most likely matches first. This progressive approach has two major benefits for time complexity. First, it produces results early (useful for real-time “on-demand” requirements), and second, it enables threshold-based pruning of comparisons. In practice, we set a similarity threshold such that if a candidate pair’s weight falls below this threshold, it is very unlikely to be an actual match and can be skipped entirely. Because the candidates are processed from highest to lowest similarity, once we reach pairs below the threshold, we can stop further comparisons in that block or for that profile. This significantly limits the number of comparisons each profile (entity) participates in. Rather than comparing each profile with all its $C'$ potential matches, each profile will on average only be compared with a small constant number $k'$ of top-ranked candidates (until the threshold condition causes pruning). If $N$ is the number of profiles under consideration (e.g. the number of query-target entities that need to be resolved on demand), the comparison cost under PPS becomes $O(N \cdot k')$. Here $k'$ can be viewed as the average number of comparisons per profile after early pruning (in many cases, $k'$ is effectively bounded by a small value, since true duplicates for an entity tend to be limited in number or caught in the first few highest-similarity links). Moreover, as matches are found, transitive matching can further reduce comparisons: if profile A matches B, and B matches C, PPS can infer A matches C without a direct comparison, avoiding redundant checks. This transitivity (clustering duplicates as they are discovered) often reduces the effective $k'$ even more. Overall, PPS ensures that the matching phase scales roughly linearly with the number of target entities, and the constant of proportionality $k'$ is kept low by prioritization and pruning. This is a substantial improvement over a naive approach that might require comparing every candidate pair (which would be on the order of $C'$ or worse per profile without PPS). 

\noindent \textbf{(4) Overall Time Complexity Estimation:} Combining all the above components, we can derive the overall time complexity of the FastER framework. Summing the costs of subgraph pattern matching, GDD constraint filtering, blocking graph construction, and progressive matching, we get: 

\[ 
T(n) = O\!\big(m \cdot |E_Q| \big) \;+\; O(C \cdot d) \;+\; O(C' \log C') \;+\; O(N \cdot k')~,
\] 

where each term corresponds to the stages analyzed: subgraph isomorphism on the original graph ($O(m \cdot |E_Q|)$), GDD filtering of $C$ candidates ($O(C \cdot d)$, which effectively becomes $O(C)$ and then $O(\beta C)$ after filtering), sorting and preparation of $C'$ candidate pairs ($O(C' \log C')$), and the progressive comparisons ($O(N \cdot k')$). For typical use cases, $|E_Q|$ and $d$ are small constants (small pattern and a fixed number of GDD rules), and $\beta$, $k'$ are small fractions or constants reflecting the effectiveness of filtering and pruning. Thus, this complexity can be simplified to: 

\[ 
O\!\Big(m \;+\; \beta C \;+\; C' \log C' \;+\; N \cdot k'\Big)~,
\] 

noting that $C' = \beta C \ll C$. In an ideal scenario (best case), the filtering is highly selective ($\beta$ is extremely small) and the similarity threshold is high enough that $k'$ remains very low. In such cases, the later terms become negligible, and the dominant cost is just scanning the relevant portion of the graph, yielding near-linear time $O(m)$ for the entire process. In a more conservative scenario (average case), we still expect $\beta$ and $k'$ to be significantly less than 1 (for instance, filtering might cut candidate pairs down to only a few percent, and each entity only needs a handful of comparisons), making the overall complexity close to linear in the size of the input graph and the number of query entities. Even in a worst-case scenario where the filtering is less effective ($\beta \approx 1$) and the threshold is low (large $k'$), the complexity $O(m + C' \log C' + N \cdot k')$ remains far more tractable than the naive $O(n^2)$ pairwise comparison of all entities. In summary, FastER’s design leverages structural constraints (GDDs), intelligent blocking, and progressive processing to achieve a much lower time complexity than brute-force entity resolution, especially benefiting from on-demand focus (small $N$) and the typically sparse nature of true duplicate links (small $k'$).

\section*{Appendix A.2: GDD Rules and Their Cypher Implementations} \label{app:A2}

\begin{figure}[!htbp]
    \centering
    \includegraphics[width=\textwidth]{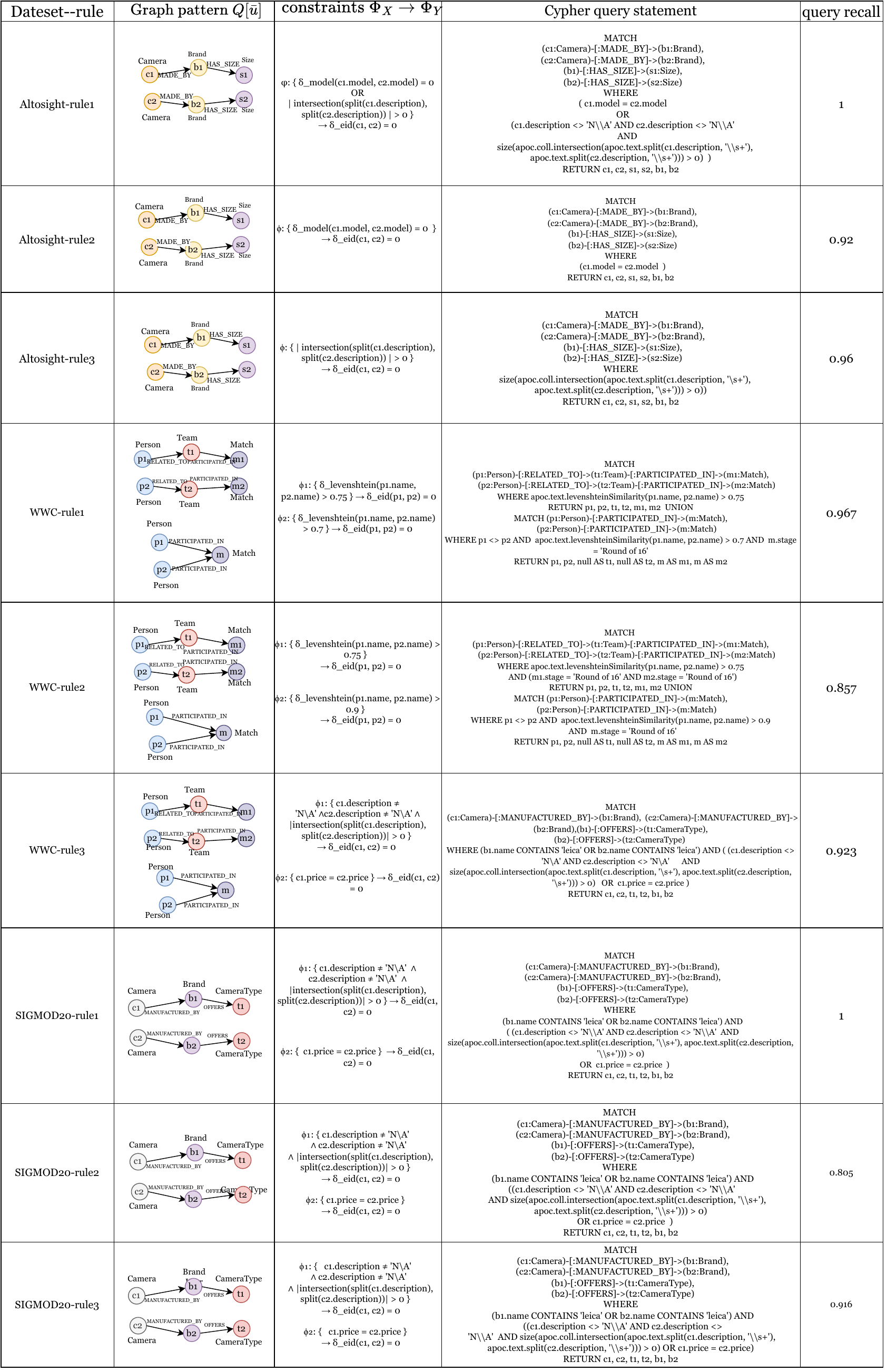}
    \caption{GDD rule examples with Cypher implementation (1).}
    \label{fig:gdd1}
\end{figure}

\begin{figure}[!htbp]
    \centering
    \includegraphics[width=\textwidth]{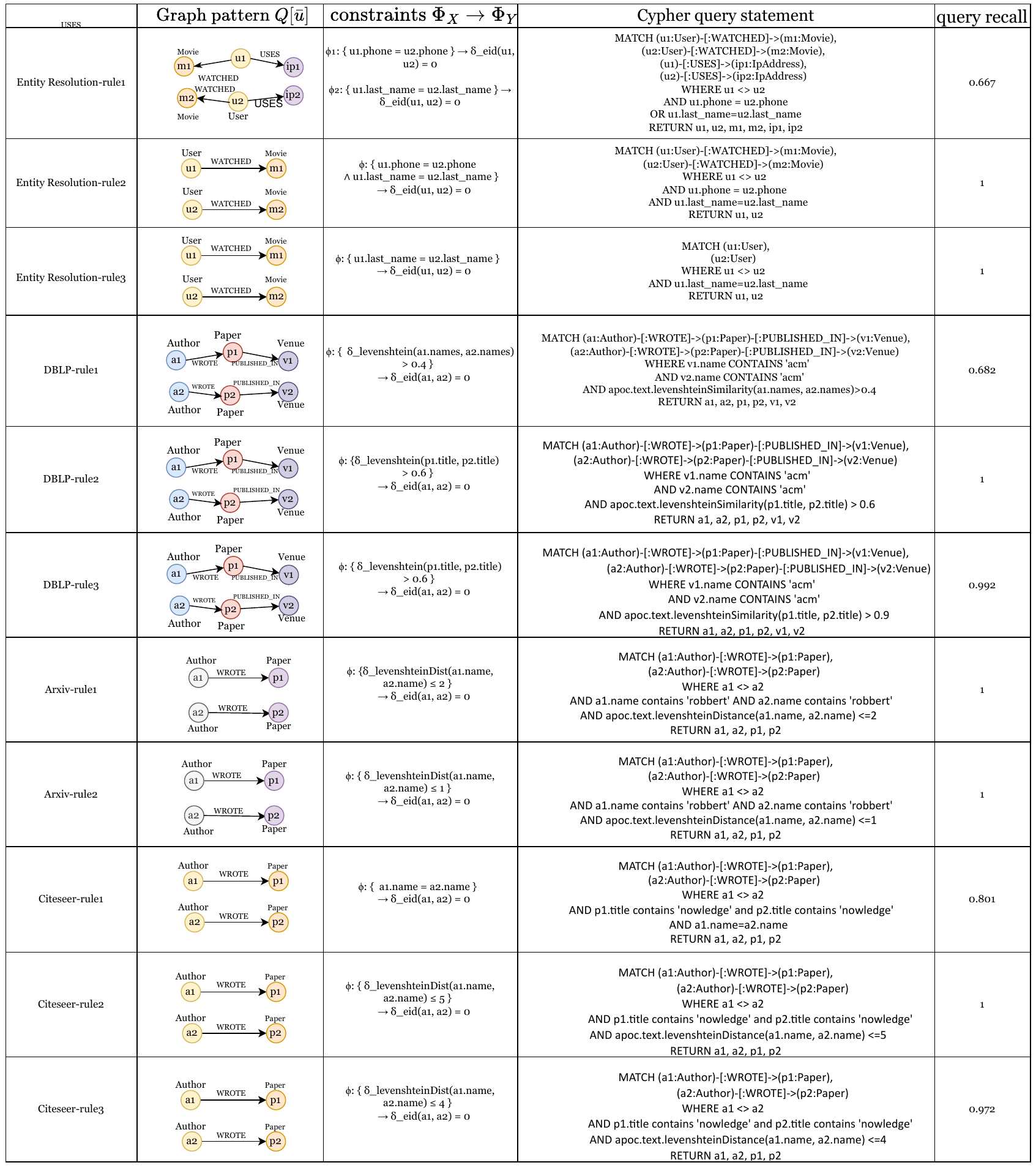}
    \caption{GDD rule examples with Cypher implementation (2).}
    \label{fig:gdd2}
\end{figure}

\begin{figure}[!htbp]
    \centering
    \includegraphics[width=\textwidth]{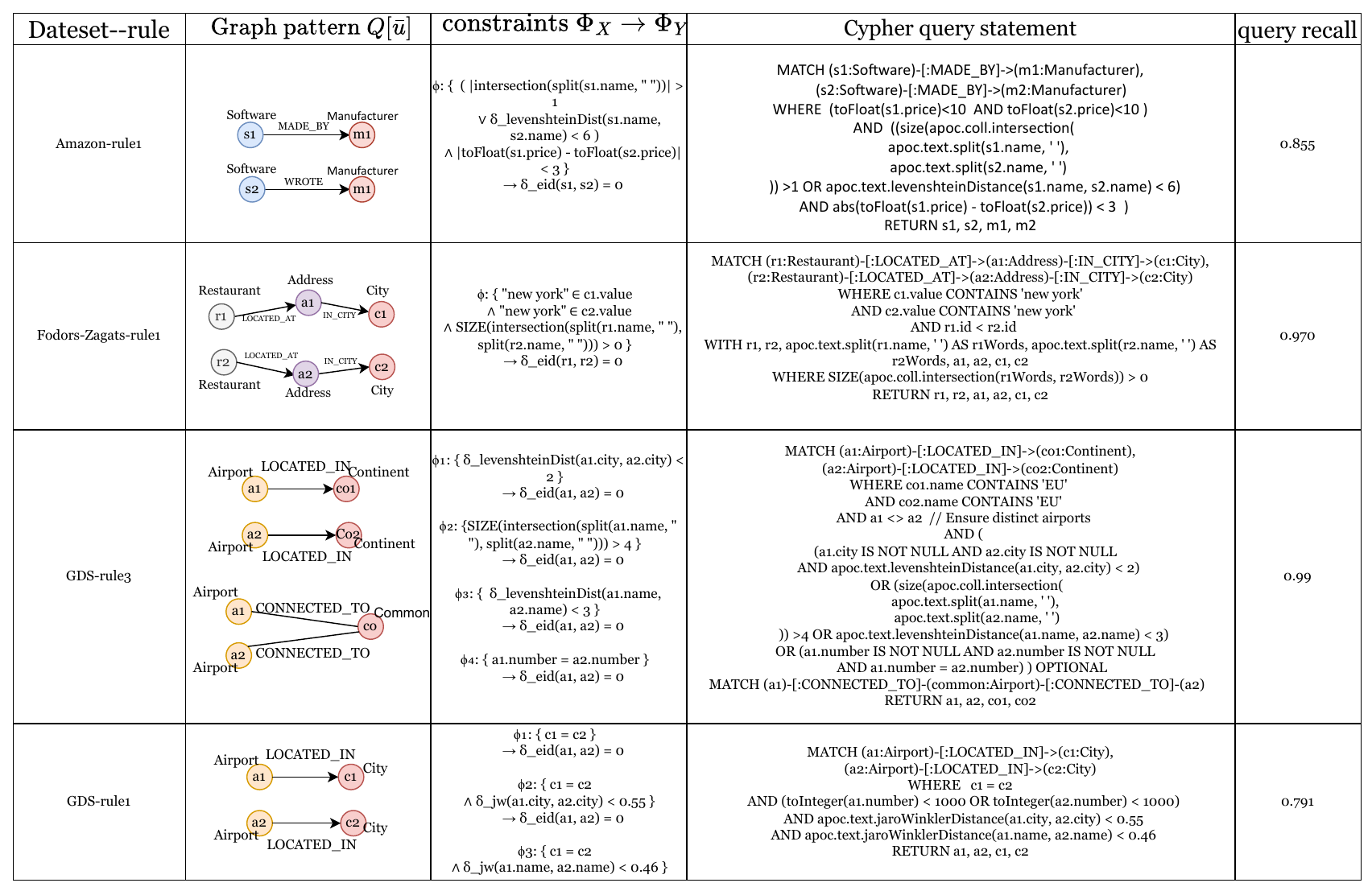}
    \caption{GDD rule examples with Cypher implementation (3).}
    \label{fig:gdd3}
\end{figure}